\documentstyle[twoside,fleqn,espcrc2,epsfig]{article}
\newcommand{\ec}{\end{center}}
\newcommand{\AmS}{{\protect\the\textfont2
  A\kern-.1667em\lower.5ex\hbox{M}\kern-.125emS}}

\def    \pp              {\ifmmode {\cal{P}} \else 
                            ${\cal{P}}$ \fi}

\def    \be             {\begin{equation}}
\def    \ee             {\end{equation}}
\def    \ba             {\begin{eqnarray}}
\def    \ea             {\end{eqnarray}}
\def    \nn             {\nonumber}
\def    \=              {\;=\;}
\def    \frac           #1#2{{#1 \over #2}}

\def    \gev            {\mbox{$\mathrm{GeV}$}}

\def    \pt             {\mbox{$p_T$}}
\def    \Q              {\ifmmode {\tiny Q} \else ${\tiny Q}$ \fi}

\def    \o              {\ifmmode {\cal{O}} \else ${\cal{Q}}$ \fi}
\def    \q              {\ifmmode {\cal{Q}} \else ${\cal{Q}}$ \fi}
\def    \oo              {\ifmmode {\cal{O}} \else 
                            $\overline{\cal{O}}$ \fi}

\def    \ups            {\ifmmode \Upsilon \else $\Upsilon$ \fi}
\def    \oneSzero       {\ifmmode {^1S_0} \else $^1S_0$ \fi}
\def    \threeSone      {\ifmmode {^3S_1} \else $^3S_1$ \fi}
\def    \onePone        {\ifmmode {^1P_1} \else $^1P_1$ \fi}
\def    \threePJ        {\ifmmode {^3P_J} \else $^3P_J$ \fi}
\def    \threePzero     {\ifmmode {^3P_0} \else $^3P_0$ \fi}
\def    \threePone      {\ifmmode {^3P_1} \else $^3P_1$ \fi}
\def    \threePtwo      {\ifmmode {^3P_2} \else $^3P_2$ \fi}

\def\oppsisx  {\mbox{$\langle  {\cal O}_1^{\psi}(^3S_1) \rangle$}}
\def\oppsihx  {\mbox{$\langle  {\cal O}_8^{\psi}(^3S_1) \rangle$}}
\def\opetahx  {\mbox{$\langle  {\cal O}_8^{\psi}(^1S_0) \rangle$}}
\def\opchizhx {\mbox{$\langle {\cal O}_8^{\psi}(^3P_0) \rangle$}}

\def    \as             {\mbox{$\alpha_s$}}

\def    \ascube         {\mbox{$\alpha_s^3$}}
\def    \asf         {\mbox{$\alpha_s^4$}}

\def \jpsi {\ifmmode {J\!/\!\psi} \else $J\!/\!\psi$ \fi}

\def \tso { \ifmmode {\langle\oo^{J/\psi}_8(\threeSone)\rangle}  \else $  \langle\oo^{J/\psi}_8(\threeSone)\rangle$\fi }

\def \delk { \ifmmode { \Delta^{J/\psi}_8(k)} \else $ \Delta^{J/\psi}_8(k)$ \fi }
\def \delt { \ifmmode { \Delta^{J/\psi}_8(3.5)} \else $ \Delta^{J/\psi}_8(3.5)$ \fi }
\def \delf { \ifmmode { \Delta^{J/\psi}_8(6.4)} \else $ \Delta^{J/\psi}_8(6.4)$ \fi }

\def\chij{\mbox{$\chi_J$}}

\def\psp {\mbox{$\psi'$}}

\def \chiz {\mbox{$\chi_{0}$}}

\def\etah {\mbox{$^1S_0^{[8]}$}}

\def\psih {\mbox{$^3S_1^{[8]}$}}
\def\psis {\mbox{$^3S_1^{[1]}$}}

\def\chijh {\mbox{$^3P_J^{[8]}$}}
\def    \rd             {{\mathrm d}} 
\def \cc {\mbox{$c \overline c$}}

\def\slash#1{{#1\!\!\!/}}

\def    \hu      {\ifmmode {h_1} \else $h_1$ \fi}
\def    \hd      {\ifmmode {h_2} \else $h_2$ \fi}
\def    \htr      {\ifmmode {h_3} \else $h_3$ \fi}
\def    \hq      {\ifmmode {h_4} \else $h_4$ \fi}

\def    \gu      {\ifmmode {g_1} \else $g_1$ \fi}
\def    \gd      {\ifmmode {g_2} \else $g_2$ \fi}
\def    \gt      {\ifmmode {g_3} \else $g_3$ \fi}
\def    \gq      {\ifmmode {g_4} \else $g_4$ \fi}

\def    \ku      {\ifmmode {k_1} \else $k_1$ \fi}
\def    \kd      {\ifmmode {k_2} \else $k_2$ \fi}
\def    \kt      {\ifmmode {k_3} \else $k_3$ \fi}
\def    \kq      {\ifmmode {k_4} \else $k_4$ \fi}

\def    \muu      {\ifmmode {\mu_1} \else $\mu_1$ \fi}
\def    \mud      {\ifmmode {\mu_2} \else $\mu_2$ \fi}
\def    \mut      {\ifmmode {\mu_3} \else $\mu_3$ \fi}
\def    \muq      {\ifmmode {\mu_4} \else $\mu_4$ \fi}

\def    \lu      {\ifmmode {\lambda^{a_1}} \else $\lambda^{a_1}$ \fi}
\def    \ld      {\ifmmode {\lambda^{a_2}} \else $\lambda^{a_2}$ \fi}
\def    \lt      {\ifmmode {\lambda^{a_3}} \else $\lambda^{a_3}$ \fi}
\def    \lq      {\ifmmode {\lambda^{a_4}} \else $\lambda^{a_4}$ \fi}
\def\tr{{\mathrm Tr}}
\def    \eps            {\ifmmode \epsilon \else $\epsilon$ \fi}
\title{$J/\psi$ Production: Tevatron and Fixed-Target Collisions}

\author{A. Petrelli\address{Argonne National Laboratory \\
        9700 S. Cass Avenue, Argonne, IL 60439 USA}%
        \thanks{Talk given at the QCD99 Euroconference, Montpellier, France, July 1999. Report No. ANL-HEP-CP-99-107.}}

\begin{document} 
\begin{abstract}
In this talk I show the results of  a fit of the NRQCD matrix elements to the 
CDF data for direct $J/\psi$  production, by including 
the radiative corrections to the $g\,g\to ^3S_1^{[1]}\, g$ 
channel and the effect of the $k_T$-smearing. Furthermore I perform the 
NLO NRQCD analysis of $J/\psi$ production in fixed-target 
proton-nucleon collisions and I fit the colour-octet matrix elements to the 
 available experimental data. The results are compared to the Tevatron ones.
\end{abstract}

\vspace{0.1in}
\maketitle
 
\section{INTRODUCTION}
The \jpsi\ production cross-section within the NRCQD factorization theory 
\cite{bbl} is given by the expression:   
\ba
\rd\sigma^\jpsi = \sum_n \langle 0\vert{\cal O}^{\jpsi}[n]\vert 0\rangle \rd\hat\sigma(\cc[n])\label{main}
\ea  
where $n=\,^{2S+1}L_J^{[1,8]}$. The relevant long-distance matrix elements 
up to order $v^4$ are $\langle\oo^{J/\psi}_8(\threeSone)\rangle$ and the 
linear combination 
$\Delta^{J/\psi}_8(k)$ = $\langle\oo^{J/\psi}_8(\oneSzero)\rangle $+$ k\, \langle\oo^{J/\psi}_8(\threePzero)\rangle/m^2$.  
The phenomenological consistency of 
the NRQCD factorization formalism 
rests upon the universality of the long-distance  matrix elements.
The phenomenological determination of the non-perturbative 
matrix elements relies on the accuracy in the computation of the short distance  kernels. In this talk I consider the Tevatron VS fixed-target 
 universality issue. First I perform a fit of \tso\ and \delt\ matrix elements to the Tevatron CDF data \cite{cdf}  by considering two possible deviation from the 
standard fits \cite{chle,cgmp,bekr}, namely a) the $O(\asf)$ 
colour-singlet contribution and b) the effect of the $k_T$-smearing.  
I successively perform a fit of the MEs to a wide fixed-target 
data sample based 
upon a NLO QCD analysis.  
\section{THE COLOUR-SINGLET $O(\asf)$ CONTRIBUTION} 
The equation (\ref{main}) is usually interpreted as a double expansion in the strong coupling $\as$ and the velocity $v$. In the \jpsi\ $p_T$-differential cross-section a third expansion parameter has to be considered, and specifically $1/p_T$ \cite{bryu}. In the triple-expansion paradigm 
It is straightforward to realize that the process 
$i\,j\,\to \threeSone^{[1]}\ k\, l$ is a priori large. 
The scaling of its partonic cross-section is in fact $O(\asf v^0/p_T^6)$ 
as compared to $O(\ascube v^4/p_T^6)$ of the C-even colour-octet 
configurations (~$\oneSzero^{[8]}$\ and $\threePJ^{[8]}$~). 
In this section I will show the effect of the channels  
$i\,j\,\to \threeSone^{[1]}\ k\, l$ in the extraction of the colour-octet
matrix elements at the Tevatron.    
The details of the calculation will appear in a forthcoming paper 
\cite{malpe}. 
In the present document I just confine myself to draw the general lines of the computation.  
The three  $O(\asf)$ channels I am going to consider are:
\ba
q\,\overline q\,\to \threeSone^{[1]}g\,g,\label{qq}\\ 
q\,g\,\to \threeSone^{[1]}q\,g,\label{qg}\\ 
g\,g\,\to \threeSone^{[1]}g\,g.\label{gg} 
\ea
\begin{figure}[htb]
\vspace{.2cm}
\centerline{\epsfig{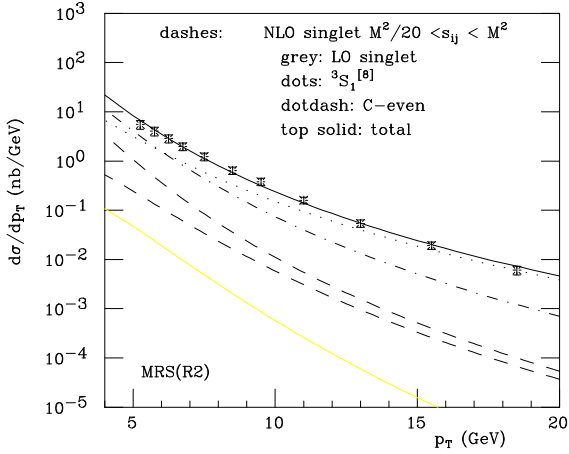}}
\caption[dummy]{\label{cut} Different channels contributing to the \jpsi\ production at the Tevatron. The NLO \psis\ channel is given by the dashed curves 
(~$s_{\rm min} = M^2_\jpsi$ (lower dash) and  $s_{\rm min} = M^2_\jpsi/20$ (upper dash)~). 
The fit of the colour-octet MEs to data are performed by considering the upper dashed curve. The resulting fitted curves are shown (~\psih\ (dots) and \delt\ (dotdash)~).}   
\end{figure}

The one-loop  colour-singlet channel is presently unknown for hadroproduction
of $\jpsi$. It is only available for \jpsi\ photoproduction \cite{kra} and annihilation into light hadrons \cite{lepmck}. 

Nevertheless, scaling arguments show that the virtual channel gives a 
subleading contribution at high $p_T$, being $O(1/p_T^8)$ its fall-off 
(~same as the born, which is already known to be negligible~). The tree-level QED-like diagrams in the channel (\ref{gg}) are also 
suppressed at high $p_T$, but they have been included to double check the global gauge 
invariance of the process. The omission of the abelian diagrams would 
generate a $1/p_T^2$-suppressed gauge dependence. To evaluate the amplitudes relative to the processes (\ref{qq})-(\ref{gg}) I 
make use of the covariant projection technique \cite{kks,gube}. 
The evaluation of the channels  (\ref{qq}) and  (\ref{qg}) is straightforward.
The  process (\ref{gg}) demands the helicity amplitude formalism: 
\ba
&&{\cal M}(P^\epsilon,\ku^\hu,\kd^\hd,\kt^\htr,\kq^\hq) =\nn\\ &&\frac{\delta_{ij}}{\sqrt{N}}\tr\left\{(\slash P+M)\gamma^\alpha {\cal A}^{\muu\mud\mut\muq}_{ij}\right\}\times\nn\\&&
{\cal E}^{\eps}_{\alpha}(P)\epsilon^\hu_\muu(\ku)\epsilon^\hd_\mud(\kd)\epsilon^\htr_\mut(\kt)\epsilon^\hq_\muq(\kq)\label{amp0}
\ea
I use the Calkul collaboration representation for the external gluons \cite{calkul} 
\ba
&&\slash\epsilon^{\pm}(k,p,q) = \frac{1}{[8(k\cdot p)(k\cdot q)(p\cdot q)]^{1/2}}
\times\;\;\;\;\;\;\;\;\;\nn\\
&&\left[ \,\slash k\;\slash p\; \slash q\,(1\mp\gamma_5) +  \slash q \;\slash p \;\slash k\,(1\pm\gamma_5) - 2(p\cdot q) \slash k\,\right]
\ea

\begin{table}[htb]
\addtolength{\arraycolsep}{0.09cm}
\renewcommand{\arraystretch}{1.4}
\begin{center}
\begin{tabular}{|l|l|l|}
\hline
&No NLO Sing &{ NLO Sing} \\
\hline
 $\langle\oo^\jpsi_8(\threeSone)\rangle$ &$1.4\pm 0.26$& ${1.5\pm 0.26}$\\
\hline
$ \Delta^\jpsi_8(3.5)$ & $12.5\pm 2.8$ & ${9.6\pm 2.8}$\\
\hline
\end{tabular}
\end{center}
\caption{\label{taba} NLO colour-singlet effect in the colour-octet MEs extraction at the Tevatron. The values are expressed in units of $10^{-2}$ GeV$^3$.
 In the first column is reported the standard fit. The second one shows the fit obtained by considering NLO \psis\ contribution with a democratic cut  on any jet pair $s_{ij}>s_{\rm min}=M^2_\jpsi/20$. The latter effect lowers the value of \delt\ but leaves \tso\ essentialy unchanged.} 
\end{table}

which corresponds to the choice of a light-like axial gauge, 
being $k$ the momentum of the given gluon and $p$, $q$ 
the two light-like reference momenta. There are three independent helicity configurations, namely $(+,+,-,-)$, 
$(+,+,+,-)$, $(+,+,+,+)$. In particular the all-plus configuration turns 
out to be zero.  Each helicity amplitude is expanded in terms of the six
colour structures represented by the six traces ${\rm tr}(\lu\ld\lt\lq)$. 
When the square is performed the C-parity symmetry allows the overall 
factorization of the colour.  

\begin{figure*}
\centerline{\epsfig{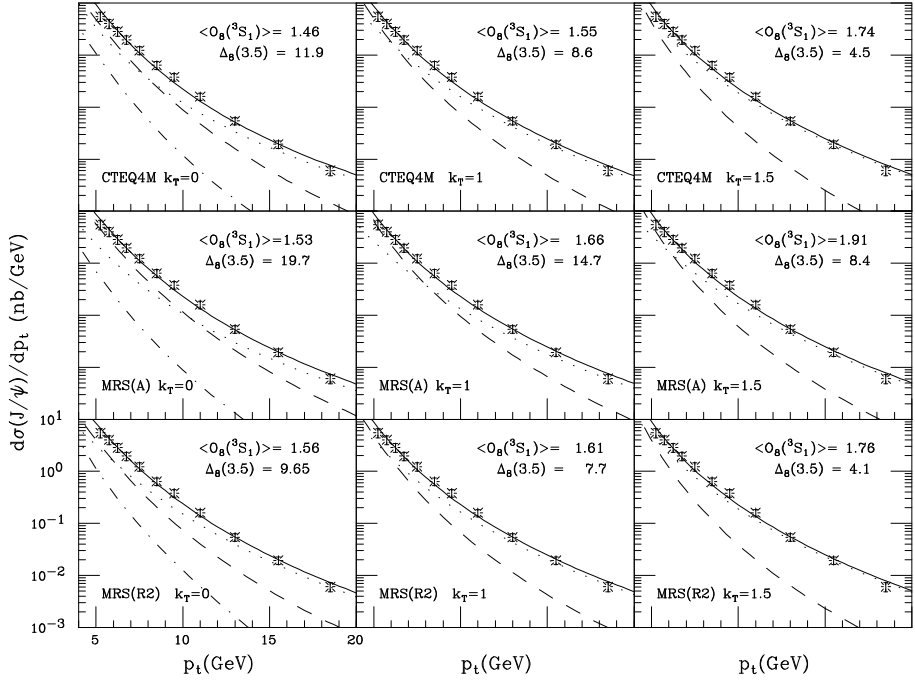}}
\caption[dummy]{\label{colla}
Different contributions to \jpsi\ production at the Tevatron. 
Dots: $\psih$. Dashes: \etah + $\chijh$. Dotdash: NLO $\psis$. The effect of 
$k_T$-smearing is included for three different values of $\langle k_T\rangle$
($\langle k_T\rangle =0,1,1.5\; \gev$). The results are given for three different sets of pdfs. The NLO \psis\ effect is only included in the $\langle k_T\rangle=0$ case.}    
\end{figure*}

\begin{table*}
\addtolength{\arraycolsep}{0.09cm}
\renewcommand{\arraystretch}{1.2}
\begin{center}
\begin{tabular}{|l l|l|l|l|}
\hline
 & & MRS(R2) & MRS(A)& CTEQ4M\\
\hline
{$\Delta^{J/\psi}_8(3.5)$}&$\langle k_T\rangle=0$&$9.6\pm2.8$&$19.7\pm3.7$&$11.9\pm 2.8$\\
              & $\langle k_T\rangle=1$&$7.7\pm2.0$&$14.8\pm2.7$&$8.6\pm2.1$\\
              &{$\langle k_T\rangle=1.5$}&{$4.1\pm1.4$}&{$8.4\pm1.9$}&{$4.5\pm1.5$}\\

\hline
{{$\langle\oo^{J/\psi}_8(\threeSone)\rangle$}}
&$\langle k_T\rangle=0$&$1.5\pm0.26$&$1.5\pm0.26$&$1.5\pm0.26$\\
                                 
&$\langle k_T\rangle=1$&$1.6\pm0.26$&$1.7\pm 0.26$&$1.5\pm0.22$\\
                                 
&$\langle k_T\rangle=1.5$&$1.7\pm0.19$&$1.9\pm0.23$&$1.7\pm0.19$\\

\hline
\end{tabular}
\end{center}
\caption{\label{tabb} Effects of intrinsic transverse momentum in the colour-octer MEs fit in \jpsi\ production at the Tevatron. Fits are performed for three different values of $\langle k_T\rangle$ and three pdfs. Values in units of $10^{-2}$ GeV$^3$.} 

\label{bovinda}
\end{table*}

\begin{figure*}
\centerline{\epsfig{file=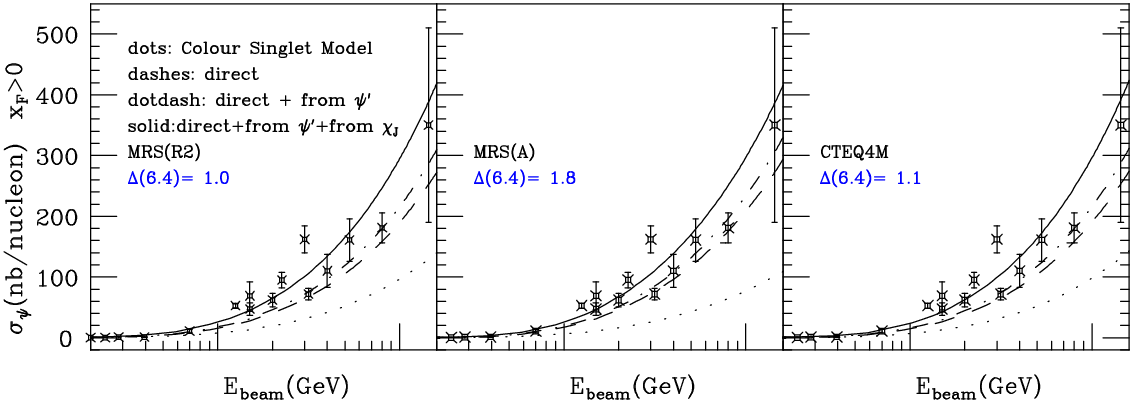,%
          height=5cm,clip=,angle=0}}
\caption[dummy]{\label{target} 
Fits of the matrix element \delf\ to the fixed-target proton-nucleon collisions data. Fits are performed for three sets of pdfs. The value of \tso\ is taken from the above Tevatron fits (~for any correspondent pdf~) with $\langle k_T\rangle=0$. (~The value of \tso\ is only sligtly affected by the intrinsic $k_T$ anyway. The indirect impact of the Tevatron $k_T$-smearing in the extraction of \delf\ is not appreciable~).}    
\end{figure*}

Being the virtual-emission channel missing, 
in performing the numerical analysis one is 
forced to put phenomenological cuts to avoid collinear and soft regions of 
the phase space. 
In particular: a) azimuth-pseudorapidity separation cut between the two final jets, b) $p_T$ cut on the two final jets.  Another way to proceed is to put a minimum invariant mass cut on any jet pair: $s_{ij}=(k_i+k_j)^2>s_{\rm min}$.
I choose the latter way. The figure (\ref{cut}) shows the cut dependence of the process for $M_\psi^2/20<s_{\rm min}<M_\psi^2$. The NLO colour-singlet contribution is given by the 
dashed lines for $s_{\rm min}=M_\psi^2$ (~lower dashes~) and  $s_{\rm min}=
M_\psi^2/20$ (~upper dashes~). The colour-singlet matrix element\footnote{For the the colour-singlet operator I use the original normalization defined in the ref. \cite{bbl}} is 
set to $\langle {\cal O}_1^{J/\psi}(^3S_1) \rangle$ = $1.2 \,\gev^3$.
As $p_T$ increases, the cut sensitivity becomes milder and milder, 
like expected. 
The fit of the colour-octet MEs is performed by assuming $s_{\rm min}= M_\psi^2/20$ and compared to the standard fit (~that is without $O(\asf)$ 
correction~). 
The results are summarized in the table  (\ref{taba}). 
The NLO colour-singlet corrections lower 
the value of the matrix element $\delt$. The VEV \tso\ instead is nailed by the high $p_T$ tail of the data distribution and is quite insensitive to the low \pt\ effects.  
At high \pt\ the $\psih$ channel develops large collinear logarithms 
$(\as \log p_T/2\,m)^n$ which make the fixed order cross section unreliable. 
The leading logarithms are resummed by using the standard DGLAP 
equation for the fragmentation function of the gluon into $\jpsi$. The accuracy of the cross section in the whole Tevatron $p_T$-range is achieved by matching 
the fixed order to the fragmentation cross-section, according to the 
following equation:
\ba
\frac{{\rd\sigma}}{\rd p_T^2}(\psih)=
\frac{\rd\sigma^{\rm{FXD}}}{\rd p_T^2}-\frac{\rd\sigma^{\rm{ASY}}}{\rd p_T^2}
+\frac{\rd\sigma^{\rm{FRG}}}{\rd p_T^2}
\ea
being
\ba
\frac{\rd\sigma^{\rm{FRG}}}{\rd p_T^2}=\frac{\rd\hat\sigma_g}{\rd p_{T,g}^2}
\otimes D_{g\to\psi}\\
\frac{\rd\sigma^{\rm{ASY}}}{\rd p_T^2}=
\left.\frac{\rd\sigma^{\rm{FXD}}}{\rd p_T^2}\right\vert_{p_T\gg m},
\ea 
where the meaning of the symbols is transparent.
From the table (\ref{taba}) can be deduced that the    
inclusion of the $O(\asf)$ \psis\ contribution in the \jpsi\
production at the Tevatron does not affect in a dramatic way the 
extraction of the colour-octet MEs. 
\begin{table}
\addtolength{\arraycolsep}{0.09cm}
\renewcommand{\arraystretch}{1.2}
\begin{center}
\begin{tabular}{|l r|l|l|l|}
\hline
&& MRS(R2) & MRS(A)& CTEQ4M\\
\hline
{{$\Delta_8^{J/\psi}(6.4)\!\!\!\!\!\!$}}
&&{$1.0$}&{$1.8$}&{$1.1$}\\
\hline
\end{tabular}
\end{center}
\caption{\label{fxtfit}Values of the fitted \delf\ at fixed-target collisions for different pdfs. In units of $10^{-2}$ GeV$^3$. } 
\end{table}

\section{THE EFFECT OF THE INTRINSIC $k_T$}
The effect of the intrinsic transverse momentum of the partons 
in the \jpsi\ differential cross-section at the Tevatron is 
phenomenologically implemented  by performing 
a gaussian smearing
of the $\pt$-distributions.  The smearing is implemented channel by channel 
for three 
values of $\langle k_T \rangle = \langle k^2_T \rangle^{1/2}$, namely
 $\langle k_T \rangle=0,1,1.5\, \gev$ and for three pdf parameterizations 
( CTEQ4M, MRS(A) and MRS(R2) ). The complete NLO calculation of colour-octet  
channels \cite{cgmmp} shows that the Sudakov effect is likely confined 
below the 2 \gev\ $\pt$-region. The region we are analysing is 
therefore free of Sudakov effects. The figure (\ref{colla}) synthetizes the 
results of the Tevatron fits with both $k_T$-smearing and colour-singlet radiative corrections. 
Since the $k_T$-smearing essentially attacks the $\pt$-slope,  
the colour-octet C-even channels are stronger affected than the flatter 
\tso\ distribution, which is instead only slightly sensitive to the 
transverse momentum of the partons. 
The basic effect of the $k_T$-kick is to tilt clockwise the dashed curves 
in figure  (\ref{colla}) and 
eventually lower the fitted value of the matrix element $\delt$.   
The obtained fits are also summarized in the table (\ref{bovinda}). 
The lack of accuracy of the NLO colour-singlet cross section at low \pt\
(~due again to the fact that the one-loop channel is still unknown~) might affect the shape 
at intermediate \pt\ once the $k_T$-smearing is turned on. That is why the NLO colour-singlet channel is only present in the $k_T=0$ case.

\section{FIXED-TARGET}
In this section I perform the fit of NRQCD MEs to a compilation of 
fixed-target data by using the NLO QCD cross sections evaluated in 
the reference \cite{cgmmp}. 
A comprehensive LO analysis can be found in the ref \cite{bero}.
The references of the experimental data 
can be found in the papers \cite{schu,bero}. 
Note that the quoted experiments do not distinguish the direct $J/\psi$ 
from the ones coming from the $\psi'$ and $\chi_J$ feed-down.
Let me fix the VEVs relative to the feed-down first. For the \chij\ feed-down 
I choose  $\langle {\cal O}_1^{\tiny\chiz}(^3P_0) \rangle/m^2$ = $4.4\times10^{-2}\, \gev^3$ and $\langle {\cal O}_8^{\tiny\chiz}(^3S_1) \rangle$ = $3.2\times10^{-3}\, \gev^3$. For the $\psi^{\prime}$:  $\langle {\cal O}_8^{\psi'}(^3S_1) \rangle$ = $4.4\times 10^{-3}\, \gev^3$ and $\Delta_8^{\psi'}(6.4)$ = $2.0\times 10^{-3} \gev^3$ (~the latter number is a result of an indepentent fit that 
will be shown somewherelse; the pdf-dependence of the $\psi'$ VEVs is not considered here since its effect on the $\jpsi$ cross section is negligible~).
Once the MEs relative to the \chij\ and \psp\ feed-down have been fixed,
I focus on the direct component of \jpsi\ production. 
 The cross section for direct \jpsi\ production 
 according to the NRQCD factorization formalism is expressed by the formula:
\ba
&&\sigma(J/\psi) =\\
&& \hat\sigma(\psis)\frac{\oppsisx}{m^5}
 +  \hat\sigma(\psih)\frac{\oppsihx}{m^5}+\nn\\&&  
{\hat\sigma(\etah)\frac{\opetahx}{m^5} + \hat\sigma(\chijh)\frac{\opchizhx}{m^7} }\nn
\ea
which is accurate up to order four in the velocity expansion.
The second line of the previous equation can be rewritten as
\ba
\sigma_{\rm }=\frac{\hat\sigma(\etah)}{m^5}\left(\opetahx + k(E_{\rm beam}) \frac{\opchizhx}{m^2}\right)\nn
\ea
The coefficient $k(E_{\rm beam})$ is independent of $E_{\rm beam}$ at LO 
(~$k(E_{\rm beam})$\,=\,7 at LO~)
and mildly dependent on $E_{\rm beam}$ at NLO: its average value is around 
6.4 ( k(100 \gev) = 6.6, k(1500 \gev) = 6.3 ).
In fixed-target collisions the matrix element $\delk$ 
appears in a linear combination which is different from the Tevatron one. On the other hand at fixed-target it is not possible to fit simultaneously 
$\delf$  and \tso\  since the all channels 
have essentially the same shape in $E_{\rm beam}$.
Therefore --following the procedure adopted in the ref \cite{bero}--
 I use the value of \tso\ fitted at Tevatron and extract \delf\
from the fixed-target data. In particular I pick  the values of  \tso\ 
obtained from the Tevatron at $k_T=0$ (~we have seen that \tso\ is not sensitive to the intrinsic $k_T$ anyway~) and I fit \delf\ for three different pdfs. The fitted curves are shown in the figure (\ref{target}). The table (\ref{fxtfit}) reports the obtained values for  $\delf$.
The NLO QCD corrections lower by about a factor of two the
 fitted value of \delf\  at fixed-target.  
 
\section{CONCLUSIONS}
Both the radiative corrections to the \psis\ channel and the $k_T$-kick 
lower the value of \delt\  extracted at the Tevatron. The previous 
effects vice versa don't have a significant impact on the 
determination of $\tso$. On the other hand the value of \delf\ obtained by fitting the fixed-target 
data is still sensibly lower than  $\delt$. If one believes that the NRQCD MEs 
are positive then $\delf>\delt$ should hold. Probably the gap would be partially bridged by the inclusion of the $O(\asf)$ 
radiative corrections also for the colour-octet channels the Tevatron. 
The large theoretical uncertainties in the evaluation of the 
charmonium total cross-section certainly affect the 
reliability of the MEs extracted from fixed-target experiments. 
Even if one does not rely in the current understanding of the 
mechanisms of charmonium production at low-$p_T$, 
the reduction of the Tevatron colour-octet MEs is still
welcome in the Hera-Tevatron universality perspective \cite{hera}.  
\section{ACKNOWLEDGEMENTS}
The results presented here have been obtained with the collaboration
 of F. Maltoni and M. L. Mangano.   
This work is supported in part by the U.S. Department
of Energy, High Energy Physics Division, Contract W-31-109-Eng-38.

\end{document}